\documentclass[lettersize,journal]{IEEEtran}
\usepackage{amsmath,amsfonts}
\usepackage{algorithmic}
\usepackage{algorithm}
\usepackage{array}
\usepackage{subfig}
\usepackage{textcomp}
\usepackage{stfloats}
\usepackage{url}
\usepackage{verbatim}
\usepackage{graphicx}
\usepackage{cite}
\usepackage{bm}
\usepackage{comment}
\usepackage{colortbl}
\usepackage{multirow}
\begin{document}

\title{EONSim: An NPU Simulator for On-Chip Memory and Embedding Vector Operations}

\author{Sangun Choi
        and Yunho Oh

\thanks{Sangun Choi and Yunho Oh are with the School of Electrical Engineering, Korea University, Seoul, South Korea. (e-mail: \{sangun\_choi, yunho\_oh\}@korea.ac.kr
).}
\thanks{Yunho Oh is the co-corresponding author.}

}

\maketitle
\begin{abstract}
Embedding vector operations are a key component of modern deep neural network workloads.  
Unlike matrix operations with deterministic access patterns, embedding vector operations exhibit input data-dependent and non-deterministic memory accesses.  
Existing neural processing unit (NPU) simulators focus on matrix computations with simple double-buffered on-chip memory systems, lacking the modeling capability for realistic embedding behavior. 
Next-generation NPUs, however, call for more flexible on-chip memory architectures that can support diverse access and management schemes required by embedding workloads.  
To enable flexible exploration and design of emerging NPU architectures, we present EONSim, an NPU simulator that holistically models both matrix and embedding vector operations.  
EONSim integrates a validated performance model for matrix computations with detailed memory simulation for embedding accesses, supporting various on-chip memory management policies.  
Validated against TPUv6e, EONSim achieves an average inference time error of 1.4\% and an average on-chip memory access count error of 2.2\%.

\end{abstract}


\section{Introduction} \label{sec:intro}

\IEEEPARstart{I}{n}  modern DNN workloads, embedding vector operations are not only an essential stage but also a dominant performance bottleneck in applications.
While existing neural processing unit (NPU) simulators primarily focus on matrix computations and fail to capture the data-dependent, non-deterministic memory behavior of embedding operations.
Unlike matrix computations with deterministic tiled accesses, embedding vector operations exhibit irregular, memory-bound access patterns that significantly affect performance.
Moreover, NPUs employ diverse on-chip memory management schemes, such as software prefetching~\cite{lee2025debunking}, software-managed caching~\cite{park2024embeddingcache}, and hardware-level cache configurations~\cite{coburn2025mtia}, which existing simulators overlook by treating on-chip memory as a simple staging buffer.

To provide detailed simulations of various on-chip memory and embedding vector operations, we propose a new NPU simulator, EONSim, based on two key insights. 
Matrix and embedding vector operations exhibit fundamentally different behaviors: the former shows deterministic, tile-based access patterns well captured by analytical models, whereas the latter involves stochastic, data-dependent accesses that require detailed cycle-level memory simulation. 
To address both challenges, EONSim integrates an analytical model for matrix operations with a fine-grained memory simulation engine for embedding vector operations, providing accurate yet efficient analysis across diverse DNN workloads. 
EONSim operates on hardware-independent embedding index traces, translating them into platform-specific memory addresses using configurable architectural parameters such as hierarchy depth, cache policy, and data layout. 
This simulation model enables realistic modeling of runtime access behavior and allows a single trace to be reused across different hardware configurations. 
EONSim further models both the vector unit and the full memory hierarchy, supporting various on-chip memory management techniques including software prefetching, cache-based management, and double buffering.

Verified against TPUv6e during DLRM inference, EONSim achieves high accuracy, reporting an average error of 1.4\% and a maximum error of 4\% in execution cycles.
EONSim also shows an average error of 2.2\% in on-chip local memory access counts.
Case studies conducted using EONSim demonstrate that memory behavior strongly affects the performance of embedding vector operations, emphasizing the importance of accurate on-chip memory modeling for architectural design.
EONSim is publicly available at GitHub~\footnote{\url{https://github.com/comsys-lab/EONSim}}.

\section{Why EONSim?} \label{sec:motivation}



Numerous NPUs have emerged to achieve better performance and energy efficiency compared to general-purpose CPUs and GPUs.
To achieve high computational throughput, NPUs typically feature multiple cores.
Each NPU core comprises dedicated compute units for scalar, vector, and matrix operations, along with a local on-chip memory.
All NPU cores share a global on-chip memory, which provides high-bandwidth data access with significantly lower latency than the off-chip memory.

While the primary acceleration target of NPUs is matrix operations, modern DNN workloads increasingly feature embedding vector operations as a critical component.
In recommendation model inference, embedding vector operations dominate the execution time, accounting for more than 90\% of total execution time~\cite{jain2023optimizing}.
The emergence of the retrieval-augmented generation (RAG) technique has also established embedding vector operations as a key stage in large language model (LLM) inference.
The retrieval stage, which involves searching and retrieving a vector database (DB) for documents related to the input query, often becomes a performance bottleneck of RAG-based inference.


\begin{figure}[t!]
	\centering
	\includegraphics[width=0.7\columnwidth]{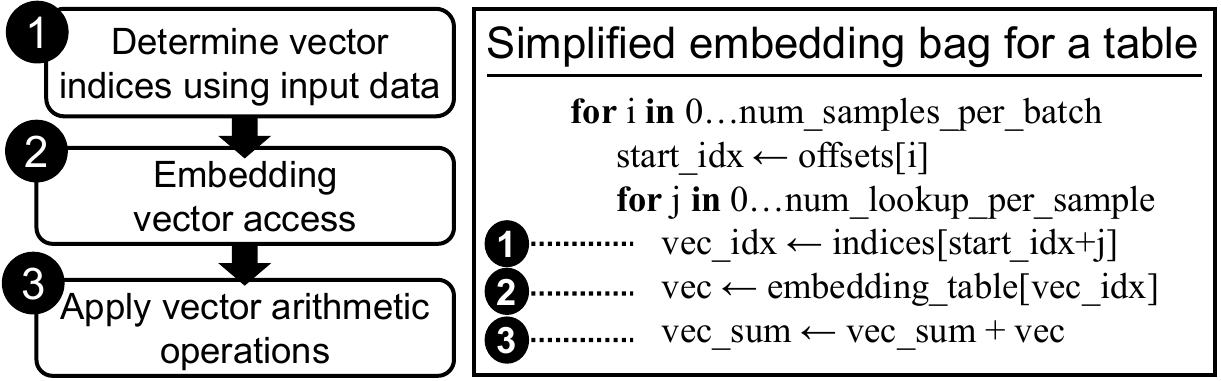}
    \caption{General process of embedding vector operations (left)
and a simplified example embedding bag operation (right).
    }
    \label{fig:evo}
\end{figure} 

While the behavior of embedding vector operations varies, embedding vector operations generally consist of three stages: (1) determining which embedding vectors to access based on input data, (2) retrieving the embedding vectors from memory, and (3) performing vector-wise arithmetic operations on the retrieved vectors.
Figure~\ref{fig:evo} shows an example of embedding bag operation in recommendation models.
Given offset values for each input sample in a batch, the NPU looks up the corresponding embedding vector indices, fetches the vectors from the embedding table, and performs summation.

Embedding vector operations exhibit distinct computation and memory access patterns that set them apart from matrix operations.
In matrix operations, statically determined dimensions of input and weight matrices primarily affect the computational workload and access patterns.
An NPU accesses every row or column of the operand matrices through the deterministic access patterns for matrix multiplication, and every element in a matrix shows the identical access counts.
In embedding vector operations, for each inference request, an NPU accesses only a small fraction of the total embedding vectors (e.g., $<$0.1\%) based on input-dependent indices that are determined at runtime.
Also, embedding vector operations exhibit highly skewed memory access patterns.
In real-world inference scenarios, certain items or tokens often appear disproportionately due to factors such as user behavior or content popularity.
Such a skewed input distribution causes an NPU to frequently access identical embedding vectors while processing multiple requests.


The significance and unique characteristics of embedding vector operations necessitate an accurate simulation infrastructure for NPU architects.
However, existing NPU simulators only focus on matrix operations, and they do not support a comprehensive analysis of embedding vector operation simulations due to two primary challenges.
%
First, existing NPU simulators cannot model input-dependent and non-deterministic access patterns of embedding vector operations.
Modeling such memory access patterns requires an NPU simulator to implement a mechanism that operates by either receiving input data to identify embedding vector access patterns or by directly receiving embedding vector access traces.
However, existing NPU simulators generate only deterministic access patterns for matrix operations~\cite{raj2025scale, ham2024onnxim, hwang2023mnpusim, yang2025pytorchsim}.
These simulators perform tiling and scheduling according to the dimensions of the matrix and the systolic array, subsequently generating a sequence of memory accesses for the matrix tiles.
This process operates independently of input data and is unsuitable for modeling the access patterns of embedding vector operations.


%

%
Second, existing NPU simulators do not support a comprehensive simulation model of various on-chip memory management techniques for embedding vector operations.
To exploit the input-dependent and skewed access patterns of embedding vector operations, NPUs utilize various on-chip management techniques, such as software embedding cache, or configuring on-chip memory as a cache~\cite{park2024embeddingcache, coburn2025mtia}.
A detailed on-chip memory model incorporating such management techniques is essential to analyze NPU behavior for embedding vector operations.
However, existing simulators assume on-chip memory as an intermediate buffer for matrix tiles.
These simulators only support a double-buffering scheme that sequentially prefetches matrix tiles, precluding an analysis of various on-chip memory management techniques~\cite{raj2025scale, hwang2023mnpusim, zhang2024llmcompass, ham2024onnxim}.
These challenges call for a novel simulation framework to analyze the architectural impact of executing embedding vector operations in NPUs.

\section{EONSim} \label{sec:implementation}


We propose EONSim, an NPU simulator that supports both matrix and embedding vector operations in DNN workloads. For matrix operations, EONSim employs an analytical model leveraging deterministic, tile-based computation and memory access.  
For embedding vector operations, we implement detailed cycle-level modeling to capture non-deterministic, data-dependent access behavior, extending simulation coverage beyond existing NPU simulators.  

Embedding vector traces are often difficult to collect on real NPUs.  
EONSim addresses this challenge by using hardware-agnostic index traces, whose patterns depend on workloads and input data rather than hardware.
During simulation, EONSim converts these traces into hardware-specific addresses, enabling trace reuse across diverse architectures.

\begin{figure}[t!]
	\centering
	\includegraphics[width=0.8\columnwidth]{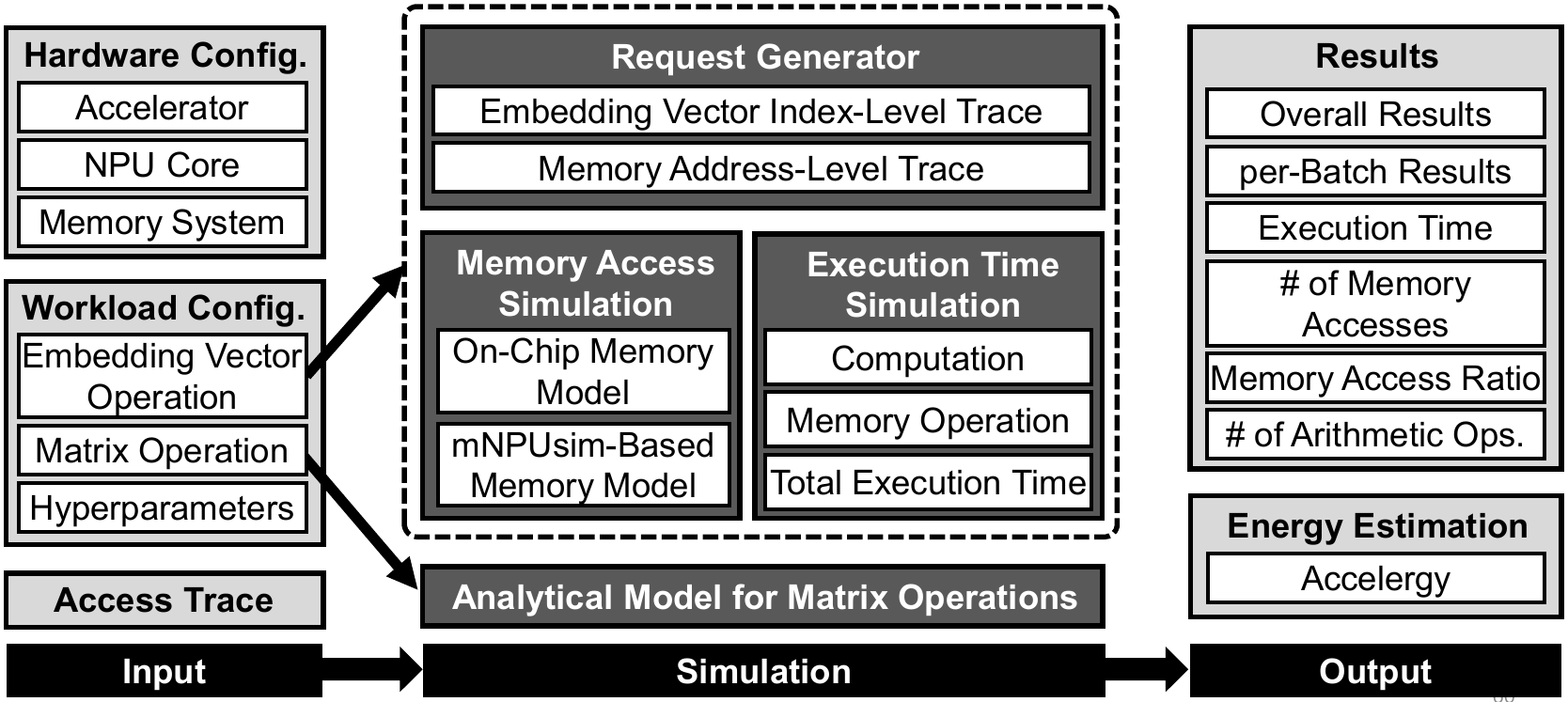}
    \caption{EONSim overview. 
    }
    \label{fig:eonsim}
\end{figure}

Figure \ref{fig:eonsim} presents an overview of EONSim, which comprises three parts: input, simulation, and output.
A comprehensive simulation requires enhanced modeling of memory hierarchy, encompassing various on-chip memory management techniques beyond the prefetching methods tailored for matrix operations.
We implement a detailed on-chip memory hierarchy and modularized on-chip memory management policies to satisfy such requirements.

\textbf{Simulation input:} EONSim requires three categories of input data for its operation.
The hardware configuration defines accelerator-level parameters such as clock frequency, the number of NPU cores, and the memory hierarchy.
Core settings detail the configuration of vector and matrix units within each core.
The memory system parameters specify memory capacity, latency, bandwidth, and access granularity.
For on-chip memory systems, users specify management policies, such as baseline double buffering, cache-based replacement policies (e.g., LRU, SRRIP), and a pinning policy to pin specific data into on-chip memory.

The workload configuration describes the computational tasks.
For matrix operations, EONSim utilizes a generalized MNK format, representing a multiplication between an $M\times K$ input matrix and an $N\times K$ weight matrix.
As this format is compatible with many NPU simulators~\cite{raj2025scale, zhang2024llmcompass, hwang2023mnpusim}, EONSim supports existing DNN model description files for matrix operations.
For embedding vector operations, users specify parameters such as the vector dimension, the number of embedding tables, and vector operations in the workload.
The configuration also defines hyperparameters, including the number of batches and the batch size.
Finally, as modeling realistic access patterns for embedding vector operations requires an access trace, EONSim takes a sequence of embedding vector indices for an embedding table.

\textbf{Simulation flow:}
To achieve fast and accurate simulation, EONSim employs distinct approaches for matrix and embedding vector operations.
For matrix operations, EONSim integrates an analytical performance model from prior work~\cite{raj2025scale, zhang2024llmcompass, park2023memory}.
This approach combines a SCALE-Sim-based model for computation cycles with an analytical model for memory operation cycles~\cite{raj2025scale, zhang2024llmcompass}.
The memory model calculates the data transfer time ($T$) using the following equation: \text{$T = D/B + L$},
where $D$ represents data size, $B$ memory bandwidth, and $L$ memory access latency.
This equation effectively models the delay of large data transfers for matrix tiles~\cite{zhang2024llmcompass, park2023memory}.

For embedding vector operations, EONSim performs a detailed memory simulation to precisely model the access patterns and their resulting performance impact.
EONSim first processes an embedding vector index-level access trace for a single table to a full access trace, based on the workload configuration (e.g., the number of embedding tables, the number of rows in a table).
EONSim then converts this index-level trace into a memory address-level access trace according to the vector dimension and memory system configuration.
In this address generation process, EONSim assumes that an NPU stores embedding vectors in consecutive virtual memory addresses.
EONSim uses the resulting memory address trace as input for the subsequent memory access simulation.

In the memory access simulation stage, the simulator identifies on-chip and off-chip memory accesses based on the provided memory access trace and the specified on-chip memory management policy.
During this process, the simulator generates intermediate access traces for both on-chip and off-chip memory accesses.
The EONSim memory model receives the intermediate access traces and performs a detailed memory access simulation.
EONSim performs the memory access simulation by adopting the off-chip memory model from a prior work~\cite{hwang2023mnpusim}, which implements an NPU memory controller and DRAMSim3-based off-chip memory modeling.

\textbf{Simulation output:}
Once the simulation is complete, EONSim outputs both overall and per-batch results.
Each result consists of various metrics, including execution time, the on-chip and off-chip memory access ratio, and the operation count for each memory and vector operation.
We integrate an Accelergy-based energy estimator into EONSim to estimate energy consumption according to the hardware configuration and operation counts~\cite{wu2019accelergy}.

\section{Evaluation} \label{sec:evaluation}

\begin{table}[t!]
\renewcommand{\arraystretch}{1.2}
\caption{
TPUv6e Hardware and DLRM Model Configuration
}
\label{table:tpuv6e}
\centering
\resizebox{0.9\columnwidth}{!}{
\begin{tabular}{|c|c|}
\hline
\textbf{Number of NPU Cores}           & 1 \\ \hline
\textbf{Systolic Array Dimension}      & 256$\times$256 \\ \hline
\textbf{Vector Unit}                   & 128 lanes, 8 sublanes \\ \hline
\textbf{Local Buffer Capacity}         & 128 MB \\ \hline
\textbf{Off-Chip Memory Capacity, BW}  & 32 GB, 1600 GB/s \\ \hline
\textbf{DLRM Model}                    & 60 embedding tables, 1M rows/table, 128-dim vectors \\ \hline
\textbf{Pooling Factor / MLP Config.}          & 120 lookups/table, 256-128-128 bottom, 128-64-1 top \\ \hline
\textbf{Eval. Parameters}              & Tables: 30–60, Batch size: 32–2048 (step 32) \\ \hline
\end{tabular}
}
\end{table}


We validate EONSim by comparing single-batch DLRM inference time with TPUv6e.
Table~\ref{table:tpuv6e} summarizes key TPUv6e parameters~\cite{TPUv6e} and DLRM model settings~\cite{jain2023optimizing}.
As TPUv6e has a single NPU core without a global buffer, it uses on-chip scratchpad memory as a temporary buffer, fetching all vectors from off-chip memory regardless of hotness.
EONSim adopts the same configuration and management policy for fair comparison.
We use the DLRM-RMC2-small model (60 embedding tables, 1M rows each, 128-dim vectors, 120 lookups per table) and vary the number of tables (30–60) and batch sizes (32–2048) to analyze scalability.

\subsection{Validation Results} \label{subsec:validation}

\begin{figure}[t!]
    \centering
    \subfloat[Varying number of embedding tables from 30 to 60]{
        \includegraphics[width=0.35\columnwidth]{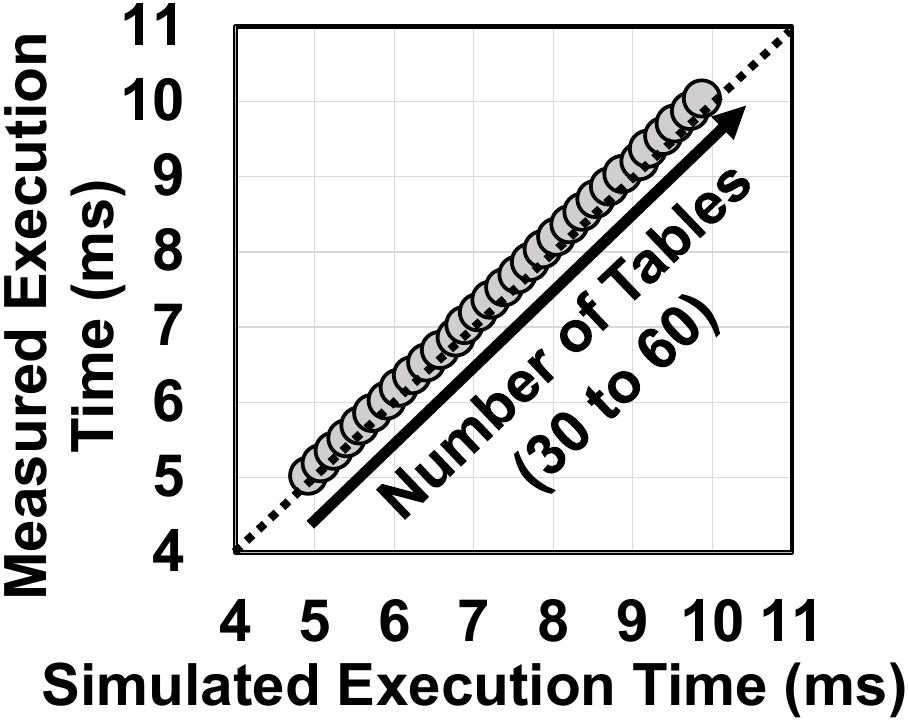}
        \label{fig:eval_val_table}
    }
    \hfill
    \subfloat[Varying batch sizes from 32 to 2048]{
        \includegraphics[width=0.35\columnwidth]{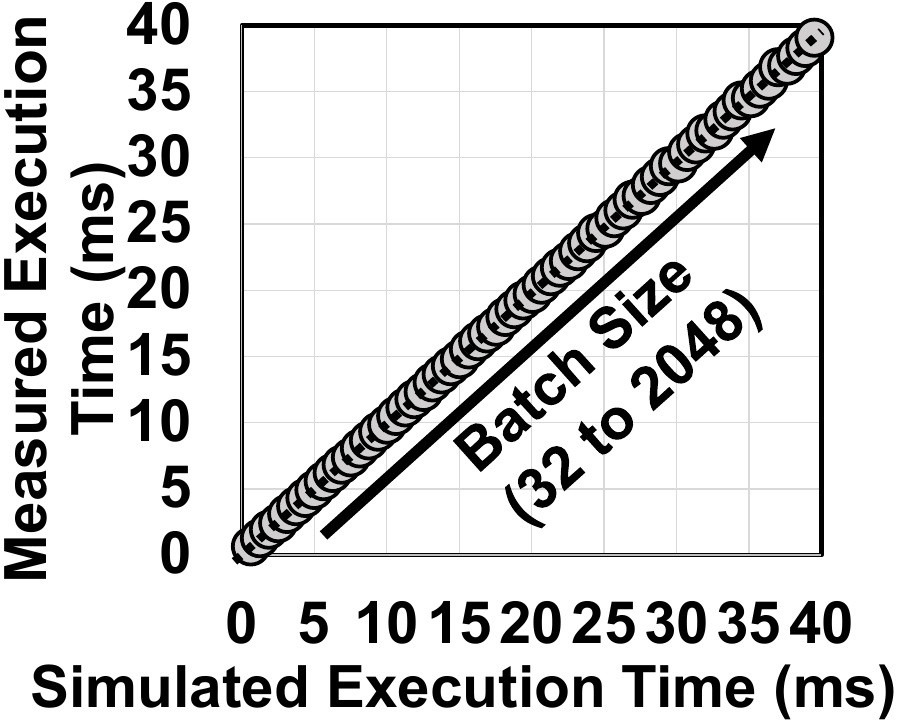}
        \label{fig:eval_val_batch}
    }
    \\[5pt] 
    \subfloat[Comparison of on-chip and off-chip memory access counts between EONSim and TPUv6e. The results in this figure are normalized to the results of TPUv6e.]{
        \includegraphics[width=0.6\columnwidth]{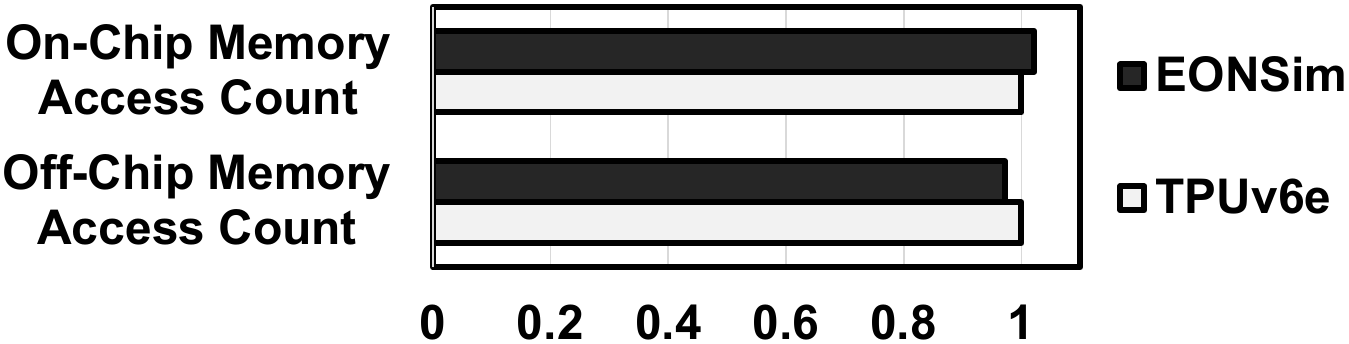}
        \label{fig:eval_val_memacc_tpu}
    }

    \caption{Validation results for EONSim across execution time and memory behavior under various configurations.}
    \label{fig:eval_val_all}
\end{figure}

\textbf{DLRM inference time and memory access behavior:}
Figure~\ref{fig:eval_val_all} shows the validation results comparing EONSim and TPUv6e.
Figures~\ref{fig:eval_val_table} and~\ref{fig:eval_val_batch} plot the measured execution time on the y-axis and the simulated execution time on the x-axis while varying the number of embedding tables and batch sizes, respectively.
EONSim achieves an average error of 2\% while varying the number of embedding tables from 30 to 60 and 1.4\% while varying the batch size from 32 to 2048.

To validate memory behavior modeling, Figure~\ref{fig:eval_val_memacc_tpu} compares on-chip and off-chip memory access counts between EONSim and TPUv6e.
As TPUv6e does not provide direct profiling support for memory access counts, we estimate them using bandwidth utilization and kernel execution time.
We compute the total data transfer per memory component and divide it by the access granularity of the TPU memory subsystem.
The resulting on-chip and off-chip memory access counts show 2.2\% and 2.8\% average error, respectively.

\textbf{On-chip memory model validation}:
Although current NPUs generally do not employ cache based on chip memory systems, future architectures may adopt such designs to improve reuse efficiency and reduce redundant memory accesses.  
EONSim provides native support for these cache like on chip management schemes, enabling simulation and analysis of potential next generation NPU configurations.  
To illustrate this capability, we evaluate a representative scenario where cache mechanisms are integrated into the NPU memory hierarchy and analyze their performance implications. 


We examine in EONSim the effect of on chip memory management techniques during recommendation model inference.  
The evaluation includes four configurations.  
SPM models the scratchpad memory used in TPU.  
LRU and SRRIP represent practical cache systems similar to the last level cache mode of MTIA.  
``Profiling" is a policy that tracks vector access frequency and pins the most frequently accessed vectors in on chip memory, up to its capacity.  
Using the TPUv6e hardware configuration and the DLRM-RMC2-small model, we measure execution time and on chip memory access ratio.  
We validate EONSim’s on-chip memory model by comparing cache behavior with ChampSim~\cite{gober2022champsim}.

\begin{figure}[t!]
    \centering
    \subfloat[Cache hit/miss comparison between EONSim and ChampSim. The results in this figure are normalized to the results of ChampSim.]{
        \includegraphics[width=0.7\columnwidth]{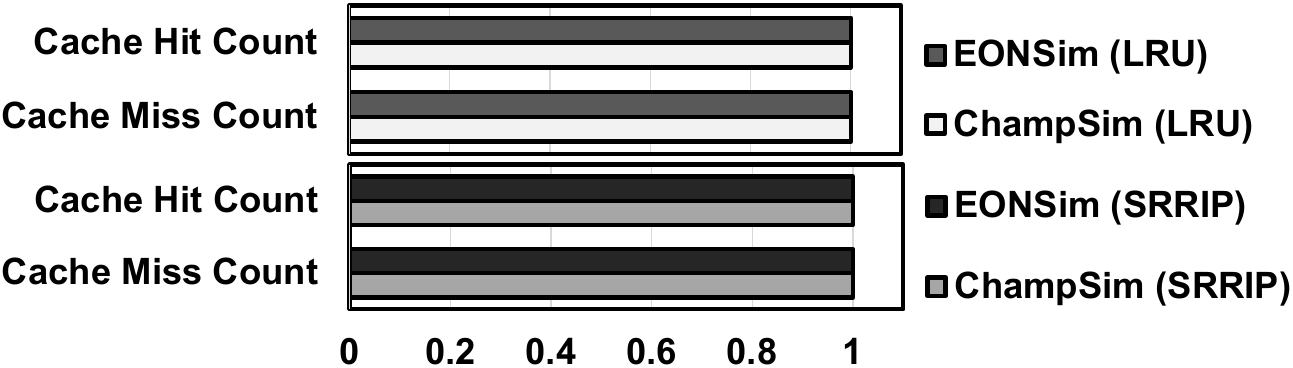}
        \label{fig:eval_val_memacc_champ}
    }
    \\[5pt] 
    \subfloat[Speedup. The results in this figure are normalized to SPM.]{
        \includegraphics[width=0.42\columnwidth]{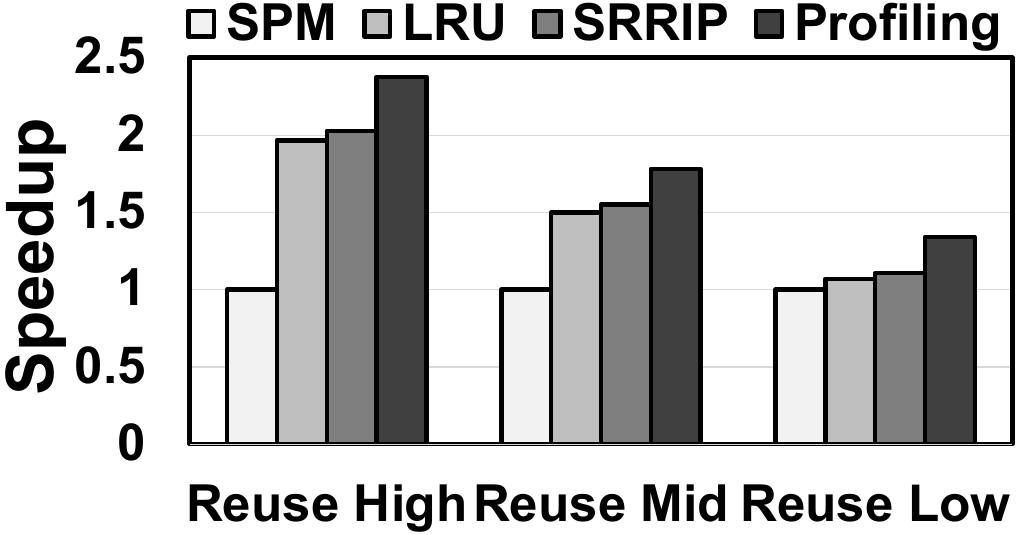}
        \label{fig:eval_dlrm_perf}
    }
    \hfill
    \subfloat[On-chip memory access ratio]{
        \includegraphics[width=0.47\columnwidth]{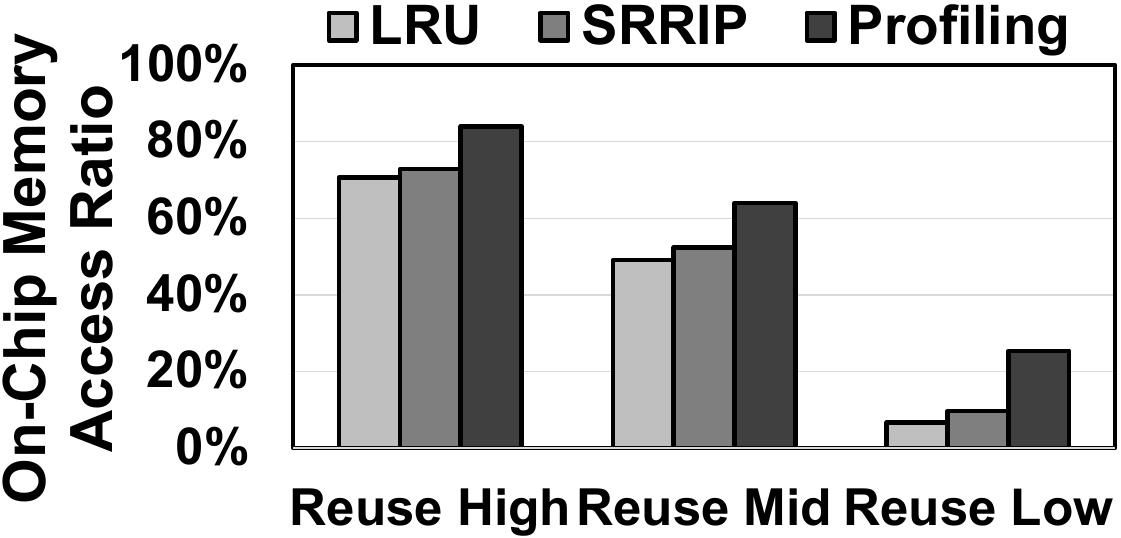}
        \label{fig:eval_dlrm_onmem}
    }    
    \caption{Performance, on-chip memory access ratio, and cache behavior comparison for recommendation model inference.}
    \label{fig:eval_dlrm}
\end{figure}

Figure~\ref{fig:eval_val_memacc_champ} compares cache hit and miss counts between EONSim and ChampSim.  
The two simulators report identical results under both LRU and SRRIP, confirming that EONSim precisely reproduces cache level behavior.  

Figure~\ref{fig:eval_dlrm_perf} presents the performance impact of different on chip management techniques.  
Cache systems with LRU and SRRIP achieve more than 1.5$\times$ speedup in the Reuse High and Mid datasets but show limited gain in Reuse Low due to frequent eviction of hot vectors.  
In Reuse High, about 4\% of vectors dominate accesses, while Reuse Low distributes them across 46\%, reducing effective reuse.  
The profiling configuration delivers the highest speedup by accurately identifying and retaining hot vectors in on chip memory.  

Figure~\ref{fig:eval_dlrm_onmem} shows the corresponding on chip memory access ratios.  
SRRIP improves the ratio by roughly 3\% over LRU, yet both remain vulnerable to cache thrashing under low access skewness.  
These observations suggest that conventional replacement policies cannot fully exploit reuse locality, whereas profiling based pinning effectively mitigates thrashing and sustains high reuse efficiency with acceptable overhead.  
From these results, we infer that next generation NPUs can benefit from hardware cache architectures or similar on chip management mechanisms that incorporate profiling or access aware policies.  
Integrating such hardware support would enable more adaptive memory behavior, improving utilization and overall performance for embedding vector workloads.

Our evaluation demonstrates that EONSim accurately models both performance and memory behavior across diverse NPU configurations.  
Beyond validation, EONSim enables detailed analysis of architectural policies through flexible configuration of on chip management schemes.  
By combining precise modeling with extensible design support, EONSim serves as a unified and reliable simulation framework for next generation NPU research.

\section{Conclusion} \label{sec:conclusion}

We present EONSim, an NPU simulator that accurately models both matrix and embedding vector operations.  
EONSim addresses the limitations of prior simulators in capturing data-dependent and non-deterministic memory behavior across hierarchical hardware structures.  
Validated against TPUv6e, EONSim achieves an average inference time error of 1.4\% and accurately reproduces on-chip and off-chip memory access counts with average errors of 2.2\% and 2.8\%, respectively.  
These results demonstrate that EONSim provides faithful timing and memory simulation for diverse DNN workloads.  



\end{document}